\documentclass[aoas,MSNbibl,nameyear,seceqn,dvips]{arximspdf}
\usepackage{algorithm}
\usepackage{algorithmic}
\usepackage{graphicx}

\doi{10.1214/10-AOAS382}
\volume{5}
\issue{1}
\pubyear{2011}
\firstpage{309}
\lastpage{336}

\makeatletter

\newtheorem{theorem}{Theorem}[section]
\newtheorem{prop}{Proposition}[section]

\makeatother

\begin{document}
\begin{frontmatter}

\title{Overlapping stochastic block models with application to the
French political blogosphere\thanksref{T1}}

\runtitle{Overlapping Stochastic Block Models}

\thankstext{T1}{Supported in part by the French Agence
Nationale de la Recherche under Grant NeMo ANR-08-BLAN-0304-01.}

\pdftitle{Overlapping stochastic block models with application to the
French political blogosphere}

\begin{aug}
\author{\fnms{Pierre} \snm{Latouche}\corref{}\ead[label=e1]{pierre.latouche@genopole.cnrs.fr}},
\author{\fnms{Etienne} \snm{Birmel\'{e}}\ead[label=e2]{etienne.birmele@genopole.cnrs.fr}}
\and
\author{\fnms{Christophe} \snm{Ambroise}\ead[label=e3]{christophe.ambroise@genopole.cnrs.fr}}

\runauthor{P. Latouche, E. Birmel\'{e} and C. Ambroise}

\affiliation{University of Evry}

\address{Laboratoire Statistique et G\'{e}nome \\
UMR CNRS 8071, INRA 1152\\
University of Evry\\
91000 Evry\\
France\\
\printead{e1} \\
\phantom{E-mail:\ }\printead*{e2}\\
\phantom{E-mail:\ }\printead*{e3}}

\end{aug}

\received{\smonth{9} \syear{2009}}
\revised{\smonth{6} \syear{2010}}

\begin{abstract}
Complex systems in nature and in society are often represented as
networks, describing the rich set of interactions between objects
of interest. Many deterministic and probabilistic clustering methods
have been developed to analyze such structures. Given a network,
almost all of them partition the vertices into \emph{disjoint}
clusters, according to their
connection profile. However, recent studies
have shown that these techniques were too restrictive and that most of the
existing networks contained overlapping clusters. To tackle this issue,
we present in this
paper the Overlapping Stochastic Block Model. Our approach allows the
vertices to belong to multiple clusters, and, to some
extent, generalizes the well-known Stochastic Block Model
[\protect\citet{ArticleNowicki2001}]. We show that the model is generically
identifiable within classes of
equivalence and we propose an approximate inference
procedure, based on global and local variational techniques. Using
toy data sets as well as the French
Political Blogosphere network and the transcriptional network of \emph
{Saccharomyces cerevisiae}, we compare our work with other approaches.

\end{abstract}



\end{frontmatter}

\setcounter{footnote}{1}

\section{Introduction} \label{secIntroduction}

Networks have been extensively studied ever since the work of
\citet{BookMoreno1934}. They are used in many scientific fields to
represent the interactions between objects of interest. For
instance, in Biology, regulatory networks can describe the regulation of
genes with transcriptional factors [\citet{ArticleMilo2002}], while
metabolic networks focus on representing
pathways of biochemical reactions [\citet{ArticleLacroix2006}]. In
the social sciences, networks are
commonly used to represent relational ties between actors [\citet
{ArticleSnijders1997}; \citet{ArticleNowicki2001}].

In this context, many deterministic and probabilistic clustering
methods have been used to acquire knowledge from the network
topology. As shown in \citet{ProceedingsNewman2007}, most of these
techniques seek specific structures in networks. Thus, some models
look for community structure where vertices are partitioned into
classes such that vertices of a class are mostly connected to vertices
of the same class [\citet{Articlehofman2008}]. They
are particularly suitable for the analysis of affiliation networks
[\citet{InbookLatouche2009}]. Most existing community discovery
algorithms are based on the modularity score of \citet
{ProceedingsGirvan2002}. However, \citet{ProceedingsBickel2009}
showed that these algorithms were (asymptotically) biased and that
using modularity scores could lead to the discovery of an incorrect
community structure, even for large graphs. The model of
\citet{ArticleHandcock2007} which extends \citet{Articlehoff2002}
is an alternative approach. Vertices are clustered depending on their
positions in a continuous latent space. They proposed a Bayesian inference
procedure, based on Markov Chain Monte Carlo (MCMC), which is
implemented in the
R package latentnet [\citet{ManualKrivitsky2009}], as well an
asymptotic BIC criterion. Other models look for disassortative mixing
in which vertices mostly connect to vertices of different
classes. They are commonly used to analyze bipartite networks
[\citet{ArticleEstrada2005}] which are present in many applications. For
more details, see \citet{ProceedingsNewman2007}.

The Stochastic Block Model
(SBM) can uncover heterogeneous structures in a large variety of
networks [\citet{InbookLatouche2009}]. Originally developed in the
social sciences, SBM is a
probabilistic generalization
[\citet{ArticleFienberg1981}; \citet{ArticleHolland1983}]
of the method described in \citet{ArticleWhite1976}. Given a network,
it assumes that each vertex belongs to a latent class among $Q$
classes and uses a $Q \times Q$ connectivity matrix $\bolds{\Pi}$ to describe
the connection probabilities [\citet{ArticleFrank1982}]. No
assumption is made on $\bolds{\Pi}$ such that SBM is a very flexible model.
In particular, it can be
used, among others, to look for community structure and disassortative
mixing. Many inference
methods have been employed to estimate the SBM parameters. They all
face the same problem. Indeed, contrary to Gaussian mixture models
or other usual mixture models, the posterior distribution $p (\mathbf
{Z}|\mathbf{X})$,
of all the hidden label variables, given the observation $\mathbf{X}$, cannot
be factorized due to conditional dependency. \citet
{ArticleNowicki2001} proposed a Bayesian
probabilistic approach. Their algorithm is implemented in the software
BLOCKS, which is part of the package StoCNET
[\citet{ManualBoer2006}]. It uses Gibbs sampling to approximate the
posterior distributions and leads to accurate {a posteriori} estimates.
Two model based criteria have been proposed to choose the optimal
value of $Q$. Thus,
\citet{ArticleDaudin2008} used an ICL criterion, based on a Laplace
approximation of the Integrated Classification Likelihood, while
\citet{InbookLatouche2009} used a nonasymptotic approximation of the
marginal likelihood. For an extensive discussion on statistical network
models and blockmodel selection, we refer to \citet{ArticleGoldenberg2009}.

A drawback of existing graph clustering techniques is that they all
partition the vertices into disjoint
clusters, while lots of objects in real world applications typically
belong to multiple groups or communities. For instance, many proteins,
so-called \emph{moonlighting proteins}, are known to have several
functions in the cells [\citet{ArticleJeffery1999}], and actors might
belong to several groups of
interests [\citet{ArticlePalla2005}]. Thus, a graph clustering method
should be able
to uncover overlapping clusters. This issue has received growing
attention in the last few years, starting
with an algorithmic approach based on small complete sub-graphs developed
by \citet{ArticlePalla2005} and implemented in the software CFinder
[\citet{ManualPalla2006}]. They defined a $k$-clique community as a
union of all $k$-cliques (complete sub-graphs of size $k$) that can be
reached from each other through a series of adjacent\footnote{Two
$k$-cliques are adjacent if they share $k-1$ vertices.}
$k$-cliques. Given a network, their algorithm first locates all cliques
and then identifies the communities using a clique--clique overlap
matrix [\citet{ArticleEverett1998}]. By construction, the resulting
communities can overlap. In order to select the optimal value of $k$, the
authors suggested a global criterion which looks for a community
structure as highly connected as possible. Small values of $k$ lead
to a giant community which smears the details of a network by merging
small communities. Conversely, when $k$ increases, the communities
tend to become
smaller, more disintegrated, but also more cohesive. Therefore, they
proposed a heuristic which consists in running their algorithm for
various values of $k$ and then to select the lowest value such that no
giant community appears.

More recent work [\citet{ArticleAiroldi2008}] proposed the
Mixed Membership Stochastic Block model (MMSB) which has been used
with success to analyze networks in many
applications [\citet{ArticleAiroldi2007}; \citet{ProceedingsAiroldi2006}].
They used variational techniques to estimate the model
parameters and proposed a criterion to select the number of
classes. As detailed in \citet{ProceedingsHeller2008}, mixed
membership models, as Latent Dirichlet Allocation [\citet
{ArticleBlei2003}], are
flexible models which can capture
partial membership [\citet
{ProceedingsGriffiths2005}; \citet{ProceedingsHeller2007}], in the form of
attribute-specific mixtures. In MMSB, a mixing weight vector
$\bolds{\pi}_{i}$ is drawn
from a Dirichlet distribution for each vertex in the network, $\pi
_{iq}$ being the probability of vertex $i$ to belong to class $q$. The
edge probability from vertex $i$ to vertex $j$ is then given by
$p_{ij}=\mathbf{Z}_{i \rightarrow j}^{\top}\mathbf{B}\mathbf
{Z}_{i \leftarrow i}$,
where $\mathbf{B}$ is a $Q \times Q$ matrix of\vspace*{1pt} connection probabilities
similar to the $\bolds{\Pi}$ matrix in SBM. The vector $\mathbf
{Z}_{i \rightarrow
j}$ is sampled from a multinomial distribution $\mathcal{M}(1,
\bolds{\pi}_{i})$ and describes the class membership of vertex
$i$ in its relation toward vertex $j$. By symmetry, the vector
$\mathbf{Z}_{i
\leftarrow j}$ is drawn from a multinomial distribution $\mathcal
{M}(1, \bolds{\pi}_{j})$ and represents the class membership of
vertex $j$ in its relation toward vertex $i$. Thus, depending on its
relations with other vertices, each vertex can belong to different
classes and, therefore, MMSB can be viewed as allowing overlapping
clusters. However, the limit of MMSB is that it does not produce edges
which are themselves influenced by the fact that some vertices belong
to multiple clusters. Indeed, for every pair $(i, j)$ of vertices, only
a single draw of $\mathbf{Z}_{i \rightarrow j}$ and $\mathbf{Z}_{i
\leftarrow j}$
determines the probability $p_{ij}$ of an edge, all the other class
memberships of vertex $i$ and $j$ toward other vertices in the network
do not play a part. In this paper we present a complementary approach
which tackles this issue.



\citet{ProceedingsFu2008} modeled overlapping clusters
on $Q$ components
by characterizing each individual $i$ by a latent $\{0,1\}$ vector
$z_i$ of
length $Q$ drawn from independent Bernoulli distributions. The $i${th}
row of the data matrix then only depends on $z_i$.
In the underlying clustering structure, $i$ belongs to the components
corresponding to a $1$ in $z_i$. Nevertheless, the proposed model needs
$Q$ parameters for each individual and supposes independence
between rows and columns of the data matrix, which is not the case when
looking for network structures.
\medskip

In this paper we propose a new model for generating
networks, depending on $(Q+1)^2+Q$ parameters, where $Q$ is the number
of components in the mixture. A latent $\{0,1\}$-vector of length $Q$
is assigned to each
vertex, drawn from products of Bernoulli distributions whose parameters
are not vertex-dependent.
Each vertex may then belong to several components, allowing overlapping
clusters, and each edge probability depends only
on the components of its endpoints.

In Section~\ref{secsbm} we briefly review the stochastic block model
introduced by \citet{ArticleNowicki2001}. In Section \ref{secmodel}
we present the overlapping stochastic block model and we show in
Section \ref{secidentifiability} that the
model is identifiable within classes of equivalence.
In Section~\ref{secinference} we propose an EM-like algorithm to
infer the parameters of the model. Finally, in Section~\ref
{secexperiments} we compare our work with other approaches using
simulated data and two real networks. We show the efficiency of our
model to detect overlapping clusters in networks.

\section{The stochastic block model}\label{secsbm}
In this paper we consider a directed binary random graph $\mathcal{G}$
represented by an $N \times N$ binary adjacency matrix $\mathbf{X}$. Each
entry $X_{ij}$ describes the presence or absence of an edge from vertex
$i$ to vertex $j$. We assume that $\mathcal{G}$ does not have any
self loop, and, therefore, the variables $X_{ii}$ will not be taken
into account. The Stochastic Block Model (SBM) introduced by \citet
{ArticleNowicki2001} associates to each vertex of a
network a latent variable $\mathbf{Z}_{i}$ drawn from a multinomial
distribution:
%
\begin{eqnarray*}
\mathbf{Z}_{i} \sim\mathcal{M}\bigl(1, \bolds{\alpha}=
(\alpha_{1}, \alpha_{2}, \dots, \alpha_{Q})\bigr),
\end{eqnarray*}
where $\bolds{\alpha}$ denotes the vector of class
proportions. As in other standard mixture models, the vector $\mathbf{Z}_{i}$
sees all its components set to zero except one such that
$Z_{iq} = 1$ if vertex $i$ belongs to class $q$. The
model then verifies
%
\begin{equation} \label{Eqconst1}
\sum_{q=1}^{Q} Z_{iq} = 1\qquad \forall i \in\{1,\dots,N\}
\end{equation}
and
%
\begin{equation} \label{Eqconst2}
\sum_{q=1}^{Q} \alpha_{q} = 1.
\end{equation}
Finally, the edges of the network are drawn from a Bernoulli distribution:
%
\begin{eqnarray*}
X_{ij} |\{Z_{iq}Z_{jl} = 1\} \sim\mathcal{B}(\pi_{ql}),
\end{eqnarray*}
where $\bolds{\Pi}$ is a $Q \times Q$ matrix of connection probabilities.
According to this model, the latent variables $\mathbf{Z}_{1}, \dots,
\mathbf{Z}_{N}$ are i.i.d. and given this latent
structure, all the edges are supposed to be independent. Note that SBM
was originally described in a more general setting, allowing any
discrete relational data. However, as explained previously, we
concentrate in the following on binary edges only.

\section{The overlapping stochastic block model}\label{secmodel}

\begin{figure}[b]

\includegraphics{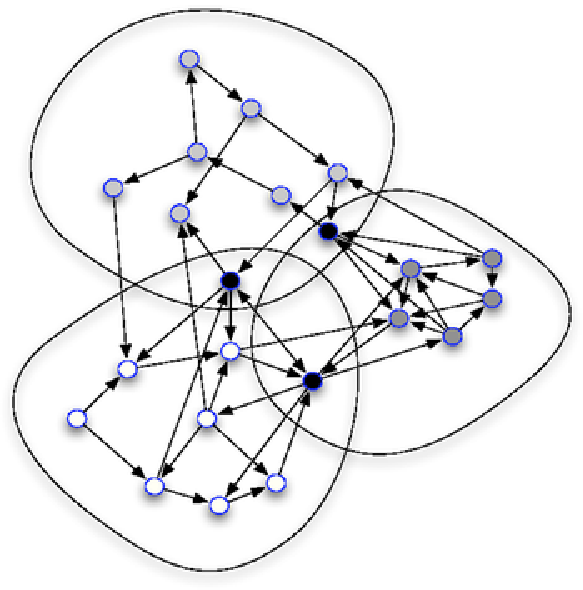}

\caption{Example of a directed graph with three overlapping clusters.}
\label{figgraphExample}
\end{figure}

In order to allow each vertex to belong to multiple classes, we relax
the constraints (\ref{Eqconst1}) and (\ref{Eqconst2}). Thus, for
each vertex $i$ of the network, we introduce a latent vector
$\mathbf{Z}_{i}$, of $Q$ independent Boolean variables $Z_{iq} \in\{
0, 1\}$,
drawn from
a multivariate Bernoulli distribution: 
%
\begin{equation} \label{EqZiDistribution}
\mathbf{Z}_{i} \sim\prod_{q=1}^{Q} \mathcal{B}(Z_{iq}; \alpha_{q})
= \prod_{q=1}^{Q} \alpha_{q}^{Z_{iq}} (1 - \alpha_{q})^{1 - Z_{iq}}.
\end{equation}
We point out that $\mathbf{Z}_{i}$ can
also have all its components set to zero which is a useful feature in
practice as described in Sections \ref{subsecoutliers} and \ref
{secexperiments}.
The edge probabilities are then given by
\begin{eqnarray*}
X_{ij}| \mathbf{Z}_{i}, \mathbf{Z}_{j} \sim\mathcal{B}( X_{ij};
\mathrm{g}(a_{\mathbf{Z}_{i}, \mathbf{Z}_{j}})) =
e^{X_{ij}a_{\mathbf{Z}_{i}, \mathbf{Z}_{j}}}\mathrm{g}(-a_{\mathbf{Z}
_{i}, \mathbf{Z}_{j}}),
\end{eqnarray*}
where
%
\begin{equation}\label{Eqaij}
a_{\mathbf{Z}_{i}, \mathbf{Z}_{j}} = \mathbf{Z}^{\top}_{i}
\mathbf{W}\mathbf{Z}_{j} +
\mathbf{Z}_{i}^{\top}\mathbf{U}+ \mathbf{V}^{\top
}\mathbf{Z}_{j} + W^{*},
\end{equation}
and $\mathrm{g}(x) = (1 + e^{-x})^{-1}$ is the logistic sigmoid
function. $\mathbf{W}$ is a $Q \times Q$ real matrix, whereas $\mathbf
{U}$ and $\mathbf{V}$ are
$Q$-dimensional real vectors. The
first term in the right-hand side of (\ref{Eqaij}) describes the
interactions between
the vertices $i$ and $j$. If $i$ belongs only to class $q$ and $j$ only
to class $l$, then only one interaction term remains
($\mathbf{Z}_{i}^{\top} \mathbf{W}\mathbf{Z}_{j} = W_{ql}$).
However, as illustrated in
Table \ref{tableaij}, the model can take more complex
interactions into account if one or both of these two vertices belong
to multiple classes (Figure \ref{figgraphExample}). Note that the
second term in
(\ref{Eqaij}) does not depend on $\mathbf{Z}_{j}$. It models the overall
capacity of vertex $i$ to connect to other vertices. By symmetry,
the third term represents the global tendency of vertex $j$ to
receive an edge. These two parameters $\mathbf{U}$ and $\mathbf{V}$
are related to
the sender/receiver effects $\delta_{i}$ and $\gamma_{j}$ in the
Latent Cluster Random Effects Model (LCREM) of \citet
{Articlekrivitsky2009}. However, contrary to LCREM, $\delta
_{i}=\mathbf{Z}
_{i}^{\top}\mathbf{U}$ and $\gamma_{j}=\mathbf{V}^{\top
}\mathbf{Z}_{j}$ depend
on the classes. In other words, two different vertices sharing the same
classes will have exactly the same sender/receiver effects, which is
not the case in LCREM. Finally, we use the scalar $W^{*}$ as a
bias, to model sparsity.

\begin{table}[t]
\tabcolsep=0pt
\caption{The values of
$a_{\mathbf{Z}_{i}, \mathbf{Z}_{j}}$ in functions of $\mathbf
{Z}_{i}$ (rows) and $\mathbf{Z}_{j}$
(columns) for an overlapping stochastic block model with $Q=2$}\label{tableaij}
\begin{tabular*}{\tablewidth}{@{\extracolsep{4in minus 4in}}lcccc@{}}
\hline
& {$\bolds{(0, 0)}$} & {$\bolds{(1, 0)}$} & {$\bolds{(0,
1)}$} & {$\bolds{(1, 1)}$}
\\
\hline
{$(0, 0)$} & $W^{*}$ & $V_{1} + W^{*}$ & $V_{2} + W^{*}$ &
$V_{1} +
V_{2} + W^{*}$
\\
{$(1, 0)$} & $U_{1} + W^{*}$ & $W_{11} +U_{1} + V_{1}+ W^{*}$ & $W_{12} + U_{1} + V_{2} + W^{*}$
&$W_{11} + W_{12} + U_{1}$\\
&&&&\qquad${}+ V_{1}+ V_{2} + W^{*}$\\
\\
{$(0, 1)$} & $U_{2} + W^{*}$ & $W_{21} + U_{2} + V_{1}+ W^{*}$
&$W_{22} + U_{2} + V_{2}+ W^{*}$ &$W_{21} +W_{22} + U_{2}$
\\
&&&&\qquad${}+ V_{1} + V_{2} + W^{*}$
\\
{$(1, 1)$} & $U_{1} + U_{2} + W^{*}$ & $W_{11} + W_{21} + U_{1}$
& $W_{12} + W_{22} + U_{1}$ & $W_{11} + W_{12} + W_{21}$
\\
&&\qquad${}+ U_{2} + V_{1} + W^{*}$& \qquad${}+ U_{2} + V_{2} + W^{*}$&\hspace*{4pt}\qquad${}+ W_{22} + U_{1} + U_{2}$
\\
&&&&\qquad${}+ V_{1} + V_{2} + W^{*}$
\\
\hline
\end{tabular*}
\end{table}

If we associate to each latent variable $\mathbf{Z}_{i}$ a vector
$\widetilde{\mathbf{Z}}_{i} = (\mathbf{Z}_{i}, 1)^{\top}$, then
(\ref{Eqaij}) can be written
%
\begin{equation} \label{EqaZiZj}
a_{\mathbf{Z}_{i}, \mathbf{Z}_{j}} = \widetilde{\mathbf
{Z}}{}_{i}^{\top}\widetilde{\mathbf{W}}\widetilde
{\mathbf{Z}}_{j},
\end{equation}
where
%
\begin{eqnarray*}
\widetilde{\mathbf{W}} =
\pmatrix{
\mathbf{W}& \mathbf{U}\cr
\mathbf{V}^{\top} & W^{*}
}
.
\end{eqnarray*}
The $\widetilde{Z}_{i(Q+1)}$'s can be seen as random variables
drawn from a Bernoulli distribution with
probability $\alpha_{Q+1}=1$. Thus, one way to think about the
model is to consider that all the vertices in the graph belong to a $(Q+1)$th
cluster which is overlapped by all the other clusters. In the
following, we will use (\ref{EqaZiZj}) to simplify the notation.

\begin{figure}[b]

\includegraphics{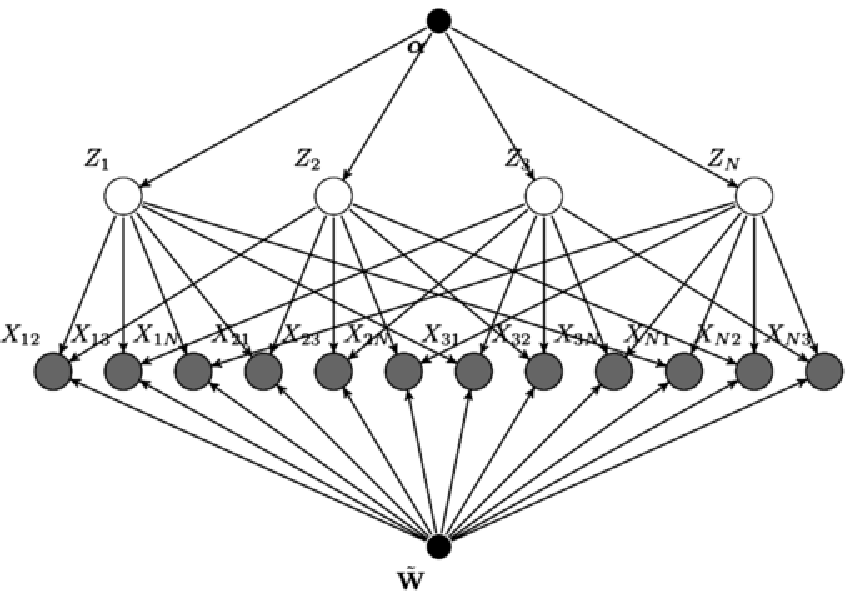}

\caption{Graphical representation of the overlapping stochastic
block model.}\label{fig2}
\end{figure}

Finally, given the latent structure $\mathbf{Z}= \{\mathbf{Z}_{1},
\dots, \mathbf{Z}_{N}\}$,
all the edges are supposed to be independent (see Figure~\ref{fig2}). Thus, when considering
directed graphs without self-loop, the Overlapping Stochastic Block
Model (OSBM) is defined through the following distributions:
%
\begin{equation} \label{EqprodBernoulli}
p(\mathbf{Z}| \bolds{\alpha}) = \prod_{i=1}^{N} \prod
_{q=1}^{Q}\alpha
_{q}^{Z_{iq}} (1 - \alpha_{q})^{1 - Z_{iq}}
\end{equation}
and
\begin{eqnarray*}
p(\mathbf{X}|\mathbf{Z}, \widetilde{\mathbf{W}}) = \prod_{i \neq
j}^{N} e^{X_{ij}a_{\mathbf{Z}_{i},
\mathbf{Z}_{j}}}{g}(-a_{\mathbf{Z}_{i}, \mathbf{Z}_{j}}).
\end{eqnarray*}

\subsection{Modeling sparsity}

As explained in \citet{ArticleAiroldi2008}, real networks are often
sparse\footnote{The corresponding adjacency matrices contain mainly
zeros.} and it is crucial to distinguish the two
sources of noninteraction. Sparsity might be the result of the rarity
of interactions in general, but it might also indicate that some
class (\emph{intra} or \emph{inter}) connection probabilities are
close to zero. For instance, social networks (see Section \ref
{subsecblog}) are often made of
communities where vertices are mostly connected to vertices of the
same community. This corresponds to classes with high \emph{intra}
connection probabilities and low
\emph{inter} connection probabilities.
In (\ref{Eqaij}) we can notice that $W^{*}$ appears in
$a_{\mathbf{Z}_{i},\mathbf{Z}_{j}}$ for every pair of vertices.
Therefore, $W^{*}$ is
a convenient parameter to model the two sources of sparsity. Indeed,
low values of $W^{*}$ result from the rarity of interactions in
general, whereas high values signify that sparsity comes from the
classes (parameters in $\mathbf{W}$, $\mathbf{U}$ and $\mathbf{V}$).

\subsection{Modeling outliers} \label{subsecoutliers}

When applied on real networks, graph clustering methods often lead to
giant classes of vertices having low output and input degrees [\citet
{ArticleDaudin2008}; \citet{InbookLatouche2009}]. These
classes are usually discarded and the analysis of networks focus on
more highly structured classes to extract useful information. The
product of Bernoulli distributions (\ref{EqprodBernoulli}) provides a
natural way to encode these ``outliers.'' Indeed, rather than using
giant classes, OSBM uses the null component such that
$\mathbf{Z}_{i}=\mathbf{0}$ if vertex $i$ is an outlier and should
not be
classified in any class.

\section{Identifiability}\label{secidentifiability}

Before looking for an optimization procedure to estimate the
model parameters, given a sample of observations (a network), it is
crucial to verify whether OSBM is identifiable. A theorem of
\citet{ArticleMatias2009} lies at the core of the results presented
in this section.

If we denote $\mathcal{F} (\Theta) = \{\mathbb{P}_{\bolds{\theta
}}, \bolds{\theta}\in
\Theta\}$, a family of models we are interested in, the classical
definition of identifiability requires that for any two different
values $\bolds{\theta}\neq\bolds{\theta}'$, the corresponding probability
distributions $\mathbb{P}_{\bolds{\theta}}$ and $\mathbb{P}_{\bolds
{\theta}'}$ are different.

\subsection{Correspondence with (nonoverlapping) stochastic block models}

Let $\Theta_{\mathrm{OSBM}}$ be the parameter space of the family of OSBMs with $Q$
classes:
\begin{eqnarray*}
\Theta_{\mathrm{OSBM}}= \bigl\{ (\bolds{\alpha}, \widetilde{\mathbf{W}}) \in
[0,1]^Q\times
\mathbb{R}^{(Q+1)^2} \bigr\}.
\end{eqnarray*}
Each $\bolds{\theta}$ in $\Theta_{\mathrm{OSBM}}$ corresponds to a random
graph model
which is defined by the distribution $p(\mathbf{X}| \bolds{\alpha},
\widetilde{\mathbf{W}})$.
The aim of this Section is to characterize whether there
exists any relation between two different parameters $\bolds{\theta}$ and
$\bolds{\theta}'$ in $\Theta_{\mathrm{OSBM}}$, leading to the same random graph
model.

We consider the (nonoverlapping) Stochastic Block
Model (SBM) introduced by \citet{ArticleNowicki2001}.
The model is defined by a set of classes $\mathcal{C}$, a vector of
class proportions $\bolds{\gamma}=\{ \gamma_{\mathbf{C}}\}_{\mathbf
{C}\in\mathcal{C}}$ verifying
$\sum_{\mathbf{C}\in\mathcal{C}}\gamma_{\mathbf{C}} =1$, and a
matrix of
connection probabilities $\bolds{\Pi}= ( \Pi_{\mathbf{C},\mathbf
{D}})_{\mathbf{C},\mathbf{D}\in\mathcal{C}^2}$.
Note that they are an infinite number of ways to represent and encode
the classes.
For simplicity, a common choice is to set
$\mathcal{C}=\{1, \dots, Q\}$ and possibly $\mathcal{C}=\{\mathbf
{C}\in\{0,
1\}^{Q}, \sum_{q=1}^{Q}C_{q}=1\}$, for a model with $Q$ classes.
The random
graphs are drawn as follows. First, the class of each vertex is sampled
from a multinomial distribution
with parameters $(1,\bolds{\gamma})$. Thus, each vertex $i$ belongs
only to
one class, and that class is $\mathbf{C}$ with probability
$\gamma_{\mathbf{C}}$. Second, the edges are drawn independently from
each other
from Bernoulli distributions, the probability of an edge $(i,j)$ being
$\Pi_{\mathbf{C},\mathbf{D}}$, if $i$ belongs to class $\mathbf{C}$
and $j$ to class~$\mathbf{D}$.

Let $\Theta_{\mathrm{SBM}}$ be the
parameter space of the family of SBMs with $2^{Q}$ classes:
\begin{eqnarray*}
\Theta_{\mathrm{SBM}}=\biggl\{ (\bolds{\gamma}, \bolds{\Pi}) \in[0,1]^{2^Q}
\times
[0,1]^{2^{2Q}}, \sum_{\mathbf{C}\in\mathcal{C}} \gamma_{\mathbf
{C}}=1\biggr\}.
\end{eqnarray*}
Considering that each possible value of the vectors $\mathbf{Z}_{i}$'s
in an
OSBM with $Q$ classes
encodes a class in a SBM with $2^{Q}$ classes (i.e., $\mathcal{C} = \{
0,1\}^{Q}$) yields a natural function:
\[
\phi\dvtx
\begin{array}{rcl}
\Theta_{\mathrm{OSBM}} & \rightarrow& \Theta_{\mathrm{SBM}} \\
(\bolds{\alpha},\widetilde{\mathbf{W}}) & \rightarrow& (\bolds{\gamma
}, \bolds{\Pi})
\end{array}
,
\]
where
\begin{eqnarray*}
\gamma_{\mathbf{C}}= \prod_{q=1}^Q \alpha_q^{C_q} (1-\alpha_q)^{1
- C_q}\qquad
\forall\mathbf{C}\in\{0, 1\}^{Q},
\end{eqnarray*}
and
%
\begin{eqnarray}
&&\Pi_{\mathbf{C},\mathbf{D}} = g( \mathbf{C}^{\top} \mathbf
{W}\mathbf{D}+
\mathbf{C}^{\top}\mathbf{U}+ \mathbf{V}^{\top}\mathbf
{D}+ W^{*} )\nonumber
\\
 &&\eqntext{\forall(\mathbf{C},\mathbf{D})
\in\{0,1\}^Q\times\{0,1\}^Q.}
\end{eqnarray}

Let $\mathcal{G}_N$ denote the set of probability measures on the
graphs of $N$ vertices.
The OSBM of parameter $\bolds{\theta}$ in $\Theta_{\mathrm{OSBM}}$ and the
SBM of parameter
$\phi(\bolds{\theta})$ in $\Theta_{\mathrm{SBM}}$ clearly induce the same measure
$\mu$
in $\mathcal{G}_N$. Thus, denoting by $\psi(\bolds{\gamma},\bolds
{\Pi})$
the probability measure in $\mathcal{G}_N$ induced by the SBM of
parameter $(\bolds{\gamma},\bolds{\Pi})$,
the problem of identifiability is to characterize
the relations between parameters $\bolds{\theta}\in\Theta_{\mathrm{OSBM}}$ and
$\bolds{\theta}' \in\Theta_{\mathrm{OSBM}}$
such that $\psi(\phi(\bolds{\theta}))=\psi(\phi(\bolds{\theta}'))$:
\[
\begin{array}{ccccc}
\Theta_{\mathrm{OSBM}}\hspace*{-15pt} & \rightarrow& \Theta_{\mathrm{SBM}} & \rightarrow& \mathcal
{G}_N, \\
\bolds{\theta}=(\bolds{\alpha},\widetilde{\mathbf{W}}) &
\stackrel{\phi}{\rightarrow} & (\bolds{\gamma}
,\bolds{\Pi}) & \stackrel{\psi}{\rightarrow} & \mu.
\end{array}
\]

The identifiability of SBM was studied by \citet{ArticleMatias2009},
who showed that the model is generically identifiable up to a
permutation of the classes. In other words, except in a set of
parameters which
has a null Lebesgue measure, two parameters imply the same random graph
model if and only if they differ only by the ordering of the classes.
Therefore, the main theorem of~\citet{ArticleMatias2009} implies the
following result.

\begin{theorem}\label{ThMatias}
There exists a set $\Theta_{\mathrm{SBM}}^{\mathrm{bad}} \subset\Theta_{\mathrm{SBM}} $ of null
Lebesgue measure such that, for every
$(\bolds{\gamma},\bolds{\Pi})$ and $(\bolds{\gamma}',\bolds{\Pi
}')$ not in $\Theta
_{\mathrm{SBM}}^{\mathrm{bad}}$, $\psi(\bolds{\gamma},\bolds{\Pi})=\psi(\bolds
{\gamma}',\bolds{\Pi}')$\vspace*{1pt}
if and only if there exists a function $P_{\nu}$ such that
$(\bolds{\gamma}',\bolds{\Pi}')=P_{\nu}((\bolds{\gamma},\bolds
{\Pi}))$, where:
\begin{itemize}[$\bullet$]
\item$\nu$ is a permutation on $\{0,1\}^{Q}$,
\item$\bolds{\gamma}'_{\mathbf{C}}=\bolds{\gamma}_{\nu(\mathbf
{C})}, \forall\mathbf{C}\in
\{0,1\}^{Q}$,
\item$\bolds{\Pi}'_{\mathbf{C},\mathbf{D}}=\Pi_{\nu(\mathbf
{C}),\nu(\mathbf{D})}, \forall(\mathbf{C}, \mathbf{D})
\in\{0,
1\}^{Q} \times\{0, 1\}^{Q}$.
\end{itemize}
\end{theorem}

Thus, studying the generical identifiability of the OSBM is equivalent
to characterizing the parameters of $\Theta_{\mathrm{OSBM}}$ verifying
$\phi( \bolds{\theta}')= P_{\nu}(\phi(\bolds{\theta}))$ for some
permutation
$\nu$ on $\{0,1\}^{Q}$.

\subsection{Permutations and inversions}

As in the case of the SBM, reordering the $Q$ classes of the OSBM and
doing the corresponding modification in $\bolds{\alpha}$ and $\widetilde
{\mathbf{W}}$
does not change the generative random graph model. Indeed, let $\sigma
$ be a permutation on $\{1,\dots,Q\}$ and let $P_{\sigma}$ denote the
function corresponding to the permutation $\sigma$ of the classes.
Then, $(\bolds{\alpha}',\widetilde{\mathbf{W}}')= P_{\sigma}(\bolds
{\alpha},\widetilde{\mathbf{W}})$
is defined by
\begin{eqnarray*}
\alpha_{q}'= \alpha_{\sigma(q)}\qquad \forall q \in\{1, \dots, Q\},
\end{eqnarray*}
and
\begin{eqnarray*}
\widetilde{\mathbf{W}}'_{q,l}=\widetilde{\mathbf{W}}_{\sigma(q),\sigma
(l)}\qquad \forall
(q,l) \in\{1,\dots,Q+1\}^{2}.
\end{eqnarray*}
Now, let $\nu$ be the permutation of $\{0,1\}^Q$ defined by
\begin{eqnarray*}
\nu(\mathbf{C})=\bigl(C_{\sigma(1)},\ldots,C_{\sigma(Q)}\bigr)\qquad \forall
\mathbf{C}\in\{0,1\}^Q.
\end{eqnarray*}
It is then straightforward to see that, for every parameter $\bolds
{\theta}$
in $\Theta_{\mathrm{OSBM}}$ and every permutation $\sigma$,
$\phi(P_{\sigma}(\bolds{\theta}))=P_{\nu}(\phi(\bolds{\theta
}))$, where $P_{\nu}$
is defined in Theorem~\ref{ThMatias}.

There is another family of operations in $\Theta_{\mathrm{OSBM}}$ which does
not change the generative random graph model, which we call inversions.
They correspond to exchanging the labels $0$ to $1$ and vice versa on
some of the coordinates of the $Z_i$'s. To give an intuition, consider
a parameter $\bolds{\theta}=(\bolds{\alpha},\widetilde{\mathbf{W}})$
in $\Theta_{\mathrm{OSBM}}$.
Let us generate graphs under the probability measure in $\mathcal
{G}_N$ induced by $\bolds{\theta}$ and consider only the first
coordinate of
the $Z_i$'s. If we denote by ``cluster~1'' the vertices whose $Z_i$'s
have a $1$ as first coordinate, the graph
sampling procedure consists in sampling the set ``cluster~1'' and then
drawing the edges conditionally on that information.
Note that it would be equivalent to sample the vertices which are not
in ``cluster~1'' and to draw the edges conditionally on that
information. Thus, there exists an equivalent reparametrization where
the $1$'s in the first coordinate correspond to the vertices which are
not in ``cluster~1.''
This is the parameter $\bolds{\theta}'$ obtained from $\bolds{\theta
}$ by an
inversion of the first coordinate.

Let $\mathbf{A}$ be any vector of $\{0,1\}^Q$. We define the {\em
$A$-inversion} $I_{\mathbf{A}}$ as follows:

\[
I_{\mathbf{A}}\dvtx
\begin{array}{rcl}
\Theta_{\mathrm{OSBM}} & \rightarrow& \Theta_{\mathrm{OSBM}} \\
(\bolds{\alpha},\widetilde{\mathbf{W}}) & \rightarrow& (\bolds{\alpha
}', \widetilde{\mathbf{W}}' )
\end{array}
,
\]
where
\begin{eqnarray*}
\alpha'_j = \cases{
1-\alpha_j, & \quad\mbox{if } $A_j=1$, \cr
\alpha_j, & \quad\mbox{otherwise}
}
 \qquad\forall j\in\{1,\ldots, Q\},
\end{eqnarray*}
and
\begin{eqnarray*}
\widetilde{\mathbf{W}}' = \mathbf{M}_{\mathbf{A}}^{\top} \widetilde
{\mathbf{W}} \mathbf{M}_{\mathbf{A}}.
\end{eqnarray*}
The matrix $\mathbf{M}_{\mathbf{A}}$ is defined by
\begin{eqnarray*}
\mathbf{M}_{\mathbf{A}} =
\pmatrix{
I-2 \operatorname{diag}(\mathbf{A}) & \mathbf{A}\cr
0\enskip \cdots\enskip 0 & 1
}
,
\end{eqnarray*}
with $\operatorname{diag}(\mathbf{A})$ being the $Q\times Q$ diagonal matrix whose diagonal
is the vector $\mathbf{A}$.

\begin{prop}
For every $\mathbf{A}\in\{0,1\}^Q$, let $\nu$ be the permutation of
$\{0,1\}
^Q$ defined by
\[
\forall\mathbf{C}\in\{0,1\}^Q\qquad \nu(\mathbf{C})_i =
 \cases{
1-C_i, &\quad\mbox{if } $A_i=1$, \cr
C_i, &\quad\mbox{otherwise}.
}
\]
Then, for every $\bolds{\theta}$ in $\Theta_{\mathrm{OSBM}}$,
\[
\phi(I_{\mathbf{A}}(\bolds{\theta})) = P_{\nu}(\phi(\bolds
{\theta})),
\]
where $P_{\nu}$ is defined in Theorem \ref{ThMatias}.
\end{prop}

\begin{pf}
Consider $\bolds{\theta}\in\Theta_{\mathrm{OSBM}}$ and $\mathbf{A}\in\{
0,1\}^Q$ and define
$(\bolds{\gamma},\bolds{\Pi})=\phi(\bolds{\theta})$ and
$(\bolds{\gamma}',\bolds{\Pi}')=\phi(I_{\mathbf{A}}(\bolds
{\theta})).$
It is straightforward to verify that
\begin{eqnarray*}
\bolds{\gamma}'_{\mathbf{C}}=\bolds{\gamma}_{\nu(\mathbf{C})}\qquad
\forall\mathbf{C}\in\{0, 1\}^{Q}.
\end{eqnarray*}
Moreover, since $M_{\mathbf{A}}
\left({ \mathbf{C}\atop
1
}
\right) = \left(
{ \nu(\mathbf{C}) \atop
1
}
\right)$, it follows that
\begin{eqnarray*}
\bolds{\Pi}'_{\mathbf{C},\mathbf{D}} & = & g \left(
\pmatrix{ \mathbf{C}^{\top} & 1
}
 \mathbf{M}_{\mathbf{A}}^{\top} \widetilde{\mathbf{W}} \mathbf
{M}_{\mathbf{A}}
\pmatrix{ \mathbf{D}\cr 1
}
 \right)
  \\
& = & g \left(
\pmatrix{ \nu(\mathbf{C})^{\top} & 1
}
 \widetilde{\mathbf{W}}
\pmatrix{ \nu(\mathbf{D}) \cr 1
}
 \right)
 \\
& = & \bolds{\Pi}_{\nu(\mathbf{C}),\nu(\mathbf{D})}.
\end{eqnarray*}
Therefore, $ \phi(I_A(\bolds{\theta})) = P_{\nu}(\phi(\bolds
{\theta})) $ .
\end{pf}

\subsection{Identifiability}

Let us define the following equivalence relation:
\[
\bolds{\theta}\sim\bolds{\theta}'\qquad\mbox{if } \exists\sigma
, \mathbf{A}|\bolds{\theta}'= I_{\mathbf{A}}(P_{\sigma}(\bolds{\theta})).
\]
To be convinced that it is an equivalence relation, note that
\begin{eqnarray*}
I_{\mathbf{A}} \circ P_{\sigma} = P_{\sigma} \circ I_{\sigma
^{-1}(\mathbf{A})}.
\end{eqnarray*}
Consider the set of equivalence classes for the
relation $\sim$. It follows that:
\begin{itemize}
\item Two parameters in the same equivalence class induce the same
measure in $\mathcal{G}_N$.
\item Each equivalence class contains a parameter $\bolds{\theta}=
(\bolds{\alpha}
,\widetilde{\mathbf{W}})$ such that
$\alpha_1\leq\alpha_2\leq\cdots\leq\alpha_Q \leq\frac{1}{2}$.
Moreover, if the $\alpha_i$'s are all distinct and strictly lower than
$\frac{1}{2}$, there is a unique such parameter in the equivalence class.
\end{itemize}

We are now able to state our main theorem about identifiability, that
is, that the model is generically identifiable up to
the equivalence relation $\sim$.

\begin{theorem}
For every $\bolds{\alpha}\in\,]0,1[^Q$, let $\beta\in\mathbb{R}^Q$
be the
vector defined by $\beta_k= - \ln(\frac{\alpha_k}{1-\alpha_k})$,
for every $k$.\vspace*{1pt}

Define $\Theta_{\mathrm{OSBM}}^{\mathrm{bad}}$ as the set of parameters $(\bolds{\alpha}
,\widetilde{\mathbf{W}})$ such that one of the following conditions holds:
\begin{itemize}
\item there exists $1\leq k\leq Q$ such that $\alpha_k=0$ or $\alpha
_k=1$ or $\alpha_k=\frac{1}{2}$,
\item there exist $1\leq k,l\leq Q$ such that $\alpha_k=\alpha_l$,
\item there exist $\mathbf{C},\mathbf{D}\in\{0,1\}^Q\times\{0,1\}
^Q$ such that
$\sum_k \beta_k C_k= \sum_k \beta_k D_k$,
\item$\phi(\bolds{\alpha}, \widetilde{\mathbf{W}}) \in\Theta
_{\mathrm{SBM}}^{\mathrm{bad}}$, set of
null measure given by Theorem \ref{ThMatias}.
\end{itemize}

Then $\Theta_{\mathrm{OSBM}}^{\mathrm{bad}}$ has a null Lebesgue measure on $\Theta
_{\mathrm{OSBM}}$ and
\[
\forall\bolds{\theta},\bolds{\theta}' \in(\Theta_{\mathrm{OSBM}}
\setminus\Theta
_{\mathrm{OSBM}}^{\mathrm{bad}})^2 \qquad
\phi(\bolds{\theta}) = \phi(\bolds{\theta}')\quad \Longleftrightarrow\quad
\bolds{\theta}\sim\bolds{\theta}
'.
\]
\end{theorem}

\begin{pf}
$\Theta_{\mathrm{OSBM}}^{\mathrm{bad}}$ is the union of a finite number of hyperplanes
or spaces which are isomorphic to hyperplanes.
Therefore, $\mu(\Theta_{\mathrm{OSBM}}^{\mathrm{bad}})=0$.
\end{pf}

Let $\bolds{\theta}=(\bolds{\alpha},\widetilde{\mathbf{W}})$, $\bolds
{\theta}'=(\bolds{\alpha}' ,\widetilde
{\mathbf{W}}')$, $\phi(\bolds{\theta})=(\bolds{\gamma},\bolds
{\Pi})$ and
$\phi(\bolds{\theta}')=(\bolds{\gamma}' ,\bolds{\Pi}')$.
As $\phi$ is constant on each equivalence class and as $\bolds{\theta
}$ and
$\bolds{\theta}'$ are not in $\Theta_{\mathrm{OSBM}}^{\mathrm{bad}}$, we can
assume that $0<\alpha_1<\cdots<\alpha_k<\frac{1}{2}$ and $0<\alpha
'_1<\cdots<\alpha'_k<\frac{1}{2}$. Proving the theorem
is then equivalent to proving that $\bolds{\theta}=\bolds{\theta}'$.

As $\phi(\bolds{\theta})=\phi(\bolds{\theta}')$, Theorem \ref
{ThMatias} ensures
that there exists a permutation $\nu\dvtx \{0,1\}^Q \to\{0,1\}^Q$ such that
\[
\cases{
\gamma'_{\mathbf{C}}  =  \gamma_{\nu(\mathbf{C})} &\quad$\forall
\mathbf{C}$,
\cr
\Pi'_{\mathbf{C},\mathbf{D}}  = \Pi_{\psi(\mathbf{C}),\psi
(\mathbf{D})} &\quad$ \forall\mathbf{C},\mathbf{D}$.
}
\]
Then, in particular,
\begin{eqnarray}\label{identeq1}
&&\biggl\{ \prod_k \alpha_k^{C_k} (1-\alpha_k)^{1-C_k}, \mathbf{C}\in\{
0,1\}^Q \biggr\}\nonumber
\\[-8pt]\\[-8pt]
&&\qquad= \biggl\{ \prod_k (\alpha'_k)^{C_k} (1-\alpha'_k)^{1-C_k}, \mathbf{C}\in
\{0,1\}
^Q \biggr\}.\nonumber
\end{eqnarray}
The minima of those two sets as well as the second lowest values are
equal, that is,
\[
\prod_k \alpha_k = \prod_k \alpha'_k \quad\mbox{and} \quad
\biggl(\prod_{k\leq Q-1} \alpha_k \biggr)(1-\alpha_Q) = \biggl(\prod_{k\leq Q-1}
\alpha'_k \biggr)(1-\alpha'_Q).
\]
Dividing those equations term by term yields $\frac{\alpha
_Q}{1-\alpha_Q}=\frac{\alpha'_Q}{1-\alpha'_Q}$ and
finally\vspace*{-3pt} $\alpha_Q=\alpha'_Q$. Dividing all terms by
$\alpha_Q^{C_Q}(1-\alpha_Q)^{1-C_Q}$ in~(\ref{identeq1}), by induction,
it follows that
%
\begin{equation}\label{identeq2}
\bolds{\alpha}=\bolds{\alpha}'.
\end{equation}

Now, for any $\mathbf{C}\in\{0,1\}^Q$, the fact that $\bolds{\gamma
}'_{\mathbf{C}} =
\bolds{\gamma}_{\nu(\mathbf{C})}$ can be written as
\begin{eqnarray*}
\prod_k \alpha_k^{C_k} (1-\alpha_k)^{1-C_k} & = & \prod_k \alpha
_k^{\nu(C)_k} (1-\alpha_k)^{1-\nu(C)_k},
\\
\sum_k C_k \ln\biggl( \frac{\alpha_k}{1-\alpha_k}\biggr) + \sum_k \ln
(1-\alpha_k) & = & \sum_k \nu(C_k) \ln\biggl( \frac{\alpha_k}{1-\alpha
_k}\biggr) + \sum_k \ln(1-\alpha_k),
\\
\sum_k \beta_k C_k & = & \sum_k \beta_k \nu(C)_k.
\end{eqnarray*}
Since $\bolds{\theta}\notin\Theta_{\mathrm{OSBM}}^{\mathrm{bad}}$, this implies that
$\nu(\mathbf{C}
)=\mathbf{C}$. As it is true for every $\mathbf{C}$, $\nu$ is in
fact the
identity function.

Therefore, for every $\mathbf{C},\mathbf{D}$, $\Pi_{\mathbf
{C},\mathbf{D}}= \Pi'_{\mathbf{C},\mathbf{D}}$, that is,
\[
\sum_{q,l}w_{ql}c_qd_l + \sum_q u_qc_q + \sum_l v_ld_l + w^* = \sum
_{q,l}w'_{ql}c_qd_l + \sum_q u'_qc_q + \sum_l v'_ld_l + w^{\prime *}.
\]

Applying it for $\mathbf{C}=\mathbf{D}=0$ implies $W^*=W^{\prime *}$.

Applying it for $\mathbf{D}=0$ and $\mathbf{C}=\bolds{\delta}_q$,
where $\bolds{\delta}_q$ is the
vector having a $1$ on the $q${th} coordinate and $0$'s elsewhere
yields $u_q+W^*=u'_q+W^{\prime *}$ and, thus, $u_q=u'_q$.

By symmetry, $\mathbf{C}=0$ and $\mathbf{D}=\delta_l$ implies $v_l=v'_l$.

Finally, $\mathbf{C}=\delta_q$ and $\mathbf{D}=\delta_l$ gives
$W_{ql}=W'_{ql}$.

Thus,
\begin{equation}\label{identeq3}
\widetilde{\mathbf{W}}=\widetilde{\mathbf{W}}'.
\end{equation}
By equations~(\ref{identeq2}) and (\ref{identeq3}), we have $\bolds
{\theta}
=\bolds{\theta}'$.

\section{Statistical inference}\label{secinference}

Given a network, our aim in this section is to estimate the OSBM parameters.

The log-likelihood of the observed data set is defined through the
marginalization: $p(\mathbf{X}| \bolds{\alpha}, \widetilde{\mathbf
{W}}) = \sum_{\mathbf{Z}}p(\mathbf{X}, \mathbf{Z}|
\bolds{\alpha}, \widetilde{\mathbf{W}})$. This summation involves
$2^{NQ}$ terms and
quickly becomes intractable. To tackle this issue, the
Expectation--Maximization (EM) algorithm has been applied on many
mixture models. However, the E-step requires the calculation of the
posterior distribution $p(\mathbf{Z}| \mathbf{X}, \bolds{\alpha},
\widetilde{\mathbf{W}})$ which cannot
be factorized in the case of networks [see \citet{ArticleDaudin2008}
for more details]. In order to obtain a tractable
procedure, we present some approximations based on global and local
variational techniques.

\subsection{The $q$-transformation}

Given a distribution $q(\mathbf{Z})$, the log-likelihood of the
observed data
set can be decomposed using
the Kullback--Leibler divergence $\mathrm{KL}(\cdot \| \cdot)$:
%
\begin{equation} \label{Eqdecomp} \ln p(\mathbf{X}| \bolds{\alpha
}, \widetilde{\mathbf{W}}) =
\mathcal{L}(q; \bolds{\alpha}, \widetilde{\mathbf{W}}) +
\operatorname{KL}(q(\cdot) \| p(\cdot|
\mathbf{X},
\bolds{\alpha}, \widetilde{\mathbf{W}})),
\end{equation}
where
%
\begin{equation} \label{EqlowerBound} \mathcal{L}(q; \bolds{\alpha},
\widetilde{\mathbf{W}}) = \sum_{\mathbf{Z}} q(\mathbf{Z}) \ln\biggl\{\frac
{p(\mathbf{X}, \mathbf{Z}| \bolds{\alpha},
\widetilde{\mathbf{W}})}{q(\mathbf{Z})}\biggr\}
\end{equation}
and
%
\begin{equation} \label{EqKL} \operatorname{KL}(q(\cdot) \| p(\cdot|
\mathbf{X},
\bolds{\alpha},
\widetilde{\mathbf{W}})) = - \sum_{\mathbf{Z}} q(\mathbf{Z}) \ln\biggl\{
\frac{p(\mathbf{Z}| \mathbf{X}, \bolds{\alpha},
\widetilde{\mathbf{W}})}{q(\mathbf{Z})}\biggr\}.
\end{equation}
The maximum $\ln p(\mathbf{X}| \bolds{\alpha}, \widetilde{\mathbf
{W}})$ of the lower bound
$\mathcal{L}$ (\ref{EqlowerBound}) is reached when $q(\mathbf{Z}) =
p(\mathbf{Z}|
\mathbf{X}, \bolds{\alpha},
\widetilde{\mathbf{W}})$. Thus, if the
posterior distribution $p(\mathbf{Z}| \mathbf{X}, \bolds{\alpha},
\widetilde{\mathbf{W}})$ was tractable,
the optimizations of $\mathcal{L}$ and $\ln p(\mathbf{X}| \bolds
{\alpha}, \widetilde
{\mathbf{W}})$, with respect to $\bolds{\alpha}$ and $\widetilde
{\mathbf{W}}$, would be
equivalent. However, in the case of
networks, $p(\mathbf{Z}| \mathbf{X}, \bolds{\alpha}, \widetilde
{\mathbf{W}})$ cannot be calculated
and $\mathcal{L}$ cannot be optimized over the entire space of
$q(\mathbf{Z})$ distributions. Thus, we restrict our search to the
class of
distributions which satisfy
%
\begin{equation} \label{Eqposterior}
q(\mathbf{Z}) = \prod_{i=1}^{N}
q(\mathbf{Z}_{i}),
\end{equation}
with
\begin{eqnarray*}
q(\mathbf{Z}_{i}) = \prod_{q=1}^{Q} \mathcal{B}(Z_{iq};
\tau_{iq})
=\prod_{q=1}^{Q}\tau_{iq}^{Z_{iq}}(1-\tau_{iq})^{1-Z_{iq}}.
\end{eqnarray*}
Each $\tau_{iq}$ is a variational parameter which corresponds to the
posterior probability of node $i$ to belong to class $q$. As for the
vector $\bolds{\alpha}$, the vectors $\bolds{\tau}_{i}=\{\tau
_{i1}, \dots,
\tau_{iQ}\}$ are not constrained to lie in the $(Q-1)$-dimensional simplex.

\begin{prop}\label{proplowerbound}
[Proof in \citet{suppappA}, Appendix \textup{A}]. 
The lower bound of the observed data log-likelihood is given by
%
\begin{eqnarray} \label{EqlowerBoundProb}
\mathcal{L}(q; \bolds{\alpha}, \widetilde{\mathbf{W}}) & =& \sum_{i
\neq j}^{N}\{
X_{ij}\tilde{\bolds{\tau}}{}_{i}^{\top} \widetilde{\mathbf
{W}}\tilde{\bolds{\tau}}_{j}
+ \mathrm{E}_{\mathbf{Z}_{i}, \mathbf{Z}_{j}}[\ln\mathrm
{g}(-a_{\mathbf{Z}_{i},\mathbf{Z}_{j}})]\}\nonumber
\\
&&{}+ \sum_{i=1}^{N} \sum_{q=1}^{Q}\{\tau_{iq}
\ln\alpha_{q} + (1 -
\tau_{iq})\ln(1 - \alpha_{q})\}
\\
&&{}- \sum_{i=1}^{N} \sum_{q=1}^{Q}\{\tau_{iq}
\ln\tau_{iq} + (1 - \tau_{iq})\ln(1 - \tau_{iq})\}.\nonumber
\end{eqnarray}
\end{prop}

Unfortunately, since the logistic sigmoid function is nonlinear,\break
$\mathrm{E}_{\mathbf{Z}_{i}, \mathbf{Z}_{j}}[\ln\mathrm
{g}(-a_{\mathbf{Z}_{i}, \mathbf{Z}_{j}})]$ in (\ref
{EqlowerBoundProb}) cannot be computed
analytically. Thus, we need a second level of approximation to
optimize the lower bound of the observed data set.

\subsection{$\xi$-transformation}

\begin{prop}[{[Proof in \citet{suppappA} in Appendix A]}] %
Given a variational parameter $\xi_{ij}$, $\mathrm{E}_{\mathbf
{Z}_{i}, \mathbf{Z}
_{j}}[\ln\mathrm{g}(-a_{\mathbf{Z}_{i}, \mathbf{Z}_{j}})]$ satisfies
\begin{eqnarray}
\qquad &&\mathrm{E}_{\mathbf{Z}_{i}, \mathbf{Z}_{j}}[\ln\mathrm
{g}(-a_{\mathbf{Z}_{i},\mathbf{Z}_{j}})]\nonumber
\\[-8pt]\\[-8pt]
&&\qquad \geq\ln\mathrm{g}
(\xi_{ij}) - \frac{(\tilde{\bolds{\tau}}{}_{i}^{\top} \widetilde
{\mathbf{W}}
\tilde{\bolds{\tau}}_{j} + \xi_{ij})}{2} - \lambda(\xi
_{ij})\bigl(\mathrm{E}_{\mathbf{Z}_{i},
\mathbf{Z}_{j}}[(\widetilde{\mathbf{Z}}{}_{i}^{\top}
\widetilde{\mathbf{W}} \widetilde{\mathbf{Z}}_{j})^2] - \xi_{ij}^2\bigr).\nonumber
\end{eqnarray}
\end{prop}

Eventually, a lower bound of the first lower bound can be computed:
%
\begin{equation}
\label{eq1}
\ln p(\mathbf{X}| \bolds{\alpha}, \widetilde{\mathbf{W}}) \geq
\mathcal{L}(q; \bolds{\alpha}, \widetilde{\mathbf{W}}) \geq
\mathcal{L}(q; \bolds{\alpha}, \widetilde{\mathbf{W}}, \bolds{\xi}) ,
\end{equation}
where
\begin{eqnarray*}
\mathcal{L}(q; \bolds{\alpha}, \widetilde{\mathbf{W}}, \bolds{\xi})
&=& \sum_{i \neq
j}^{N}\biggl\{\biggl(X_{ij} - \frac{1}{2}\biggr)\tilde{\bolds{\tau
}}{}_{i}^{\top}
\widetilde{\mathbf{W}} \tilde{\bolds{\tau}}_{j} + \ln
\mathrm{g}(\xi_{ij}) - \frac{\xi_{ij}}{2}
\\
&&{}\hspace*{19pt}- \lambda(\xi_{ij})\bigl(\operatorname{Tr}(\widetilde{\mathbf
{W}}{}^{\top
}\widetilde{\mathbf{E}_{i}}\widetilde{\mathbf{W}} \bolds{\Sigma}_{j})
+\tilde{\bolds{\tau}}{}_{j}^{\top}\widetilde{\mathbf
{W}}{}^{\top}\widetilde{\mathbf{E}
_{i}}\widetilde{\mathbf{W}}\tilde{\bolds{\tau}}_{j} - \xi_{ij}^2
\bigr)\biggr\}
 \\
&&{}+ \sum_{i=1}^{N} \sum_{q=1}^{Q}\{\tau_{iq}
\ln\alpha_{q} + (1 -
\tau_{iq})\ln(1 - \alpha_{q})\}
\\
&&{}- \sum_{i=1}^{N} \sum_{q=1}^{Q}\{\tau_{iq}
\ln\tau_{iq} + (1 - \tau_{iq})\ln(1 - \tau_{iq})\}.
\end{eqnarray*}

The resulting variational EM algorithm (see Algorithm \ref
{algoc2_algo}) alternatively computes the posterior probabilities
$\bolds{\tau}_{i}$ and
the parameters $\bolds{\alpha}$ and $\widetilde{\mathbf{W}}$ maximizing
\[
\max_{\bolds{\xi}} \mathcal{L}(q; \bolds{\alpha}, \widetilde{\mathbf
{W}}, \bolds{\xi}).
\]
The optimization equations are given in \citet{suppappA}, Appendix B.

\begin{algorithm}[t]
\caption{Overlapping stochastic block model for directed graphs
without self loop.}
{\fontsize{9}{10}\selectfont
\label{algoc2_algo}
// INITIALIZATION\\
\hspace*{5pt} Initialize $\bolds{\tau}$ with an Ascendant Hierarchical
Classification algorithm
Sample $\widetilde{\mathbf{W}}$ from a zero mean\\
\hspace*{5pt}   $\sigma^{2}$ spherical Gaussian
distribution\\
// OPTIMIZATION\\
\hspace*{5pt} \textbf{repeat}\\
\hspace*{20pt} // $\xi$-transformation\\
\hspace*{20pt} \textbf{for} $(i, j)\in V$ \textbf{do}\\
\hspace*{38pt} $\xi_{ij} \leftarrow\sqrt{\operatorname{Tr}(\widetilde{\mathbf
{W}}{}^{\top}\widetilde
{\mathbf{E}_{i}}\widetilde{\mathbf{W}}
\bolds{\Sigma}_{j}) + \tilde{\bolds{\tau}}{}_{j}^{\top}\widetilde
{\mathbf{W}}{}^{\top
}\widetilde{\mathbf{E}_{i}}\widetilde{\mathbf{W}}\tilde{\bolds{\tau}}_{j}}$
\\
\hspace*{20pt} \textbf{end}
\\
\hspace*{20pt} // M-step
\\
\hspace*{20pt} \textbf{for} $q=1:Q$ \textbf{do}
\\
\hspace*{38pt} $\alpha_{q} \leftarrow\frac{\sum_{i=1}^{N} \tau_{iq}}{N}$
\\
\hspace*{20pt} \textbf{end}
\\
\hspace*{20pt} Optimize $\mathcal{L}(q; \bolds{\alpha}, \widetilde{\mathbf{W}},
\bolds{\xi})$ with
respect to $\widetilde{\mathbf{W}}$, with a gradient based
optimization algorithm\\
\hspace*{20pt}  [e.g., quasi-Newton method of \citet{ArticleBroyden1970}]
\\
\hspace*{20pt} // E-step\\
\hspace*{20pt} \textbf{repeat}\\
\hspace*{40pt}\textbf{for} $i=1:N$ \textbf{do}\\
\hspace*{52pt} Optimize $\mathcal{L}(q; \bolds{\alpha}, \widetilde{\mathbf{W}},
\bolds{\xi})$
with respect to $\bolds{\tau}_{i}$, with a box constrained ($\tau_{iq}
\in[0,1]$) \\
\hspace*{52pt}
gradient based optimization algorithm [e.g., Byrd  method \citet{ArticleByrd1995}]
\\
\hspace*{40pt}\textbf{end}
\\
\hspace*{20pt} \textbf{until} $\bolds{\tau}$ \textit{converges}
\\
\hspace*{5pt} \textbf{until} $\mathcal{L}(q; \bolds{\alpha}, \widetilde{\mathbf{W}},
\bolds{\xi})$ \textit{converges}}
\end{algorithm}

The computational cost of the algorithm is equal to $O
(N^{2}Q^{4})$. For comparison the computational cost of the methods
proposed by \citet{ArticleDaudin2008} and
\citet{InbookLatouche2009} for (nonoverlapping) SBM is equal to
$O (N^{2}Q^{2})$. Analyzing a sparse network with $100$ nodes takes
about ten seconds on a dual core, and about a minute for dense networks.

For all the experiments we present in the following section, set
$\sigma^{2}=0.5$ and we used
the Ascendant Hierarchical Classification (AHC) algorithm implemented
in the
R package ``mixer'' which is available
at the following:
\url{http://cran.r-project.org/web/packages/mixer}.

\section{Experiments}\label{secexperiments}

We present some results of the experiments we carried out to assess
OSBM. Throughout our experiments,
we compared our approach to SBM (the nonoverlapping version
of OSBM), the Mixed Membership Stochastic Block model (MMSB) of \citet
{ArticleAiroldi2008}, and the work of \citet{ArticlePalla2005},
implemented in the
software (Version 2.0.1) CFinder [\citet{ManualPalla2006}].

In order to perform inference in SBM, we used the variational Bayes
algorithm of \citet{InbookLatouche2009} which approximates the
posterior distribution over the latent variables and model parameters,
given the edges. We computed the Maximum A Posteriori (MAP) estimates
and obtained the class membership vectors $\mathbf{Z}_{i}$. We recall
that SBM
assumes that each vertex belongs to a single class and, therefore, each
vector $\mathbf{Z}_{i}$ has all its components set to zero except one, such
that $Z_{iq}=1$ if vertex $i$ is classified into class $q$. For OSBM,
we relied on the variational approximate inference procedure described
in Section \ref{secinference} and computed the MAP estimates.
Contrary to SBM, each vertex can belong to multiple clusters and,
therefore, the vectors $\mathbf{Z}_{i}$ can have multiple components
set to
one. As described in Section \ref{secIntroduction}, MMSB can also be
viewed as allowing overlapping clusters. For more details, we refer to
\citet{ArticleAiroldi2008}. In order to estimate the MMSB mixing
weight vectors $\bolds{\pi}_{i}$, we used the collapsed Gibbs
sampling approach implemented in the R package lda [\citet
{ManualChang2010}]. We then converted each vector $\bolds{\pi
}_{i}$ into a binary membership vector $\mathbf{Z}_{i}$ using a
threshold $t$.
Thus, for $\pi_{iq}\geq t$, we set $Z_{iq}=1$ and $Z_{iq}=0$
otherwise. In all the experiments we carried out, we defined $t=1/Q$
and we found that for higher values MMSB tended to behave like SBM.
Finally, we considered CFinder which is a widely used algorithmic
approach to uncover overlapping communities. As described in Section
\ref{secIntroduction}, CFinder looks for $k$-clique communities where
each $k$-clique community is a union of all $k$-cliques (complete
sub-graphs of size $k$) that can be reached from each other through a
series of adjacent $k$-cliques. The algorithm first locates all cliques
and then identifies the communities and overlaps between communities
using a clique--clique overlap matrix [\citet{ArticleEverett1998}].
Vertices that do not belong to any $k$-clique are seen as outliers and
not classified.

Contrary to OSBM (and CFinder), SBM and MMSB cannot deal with outliers.
Therefore, to obtain fair comparisons between the approaches, when OSBM
was run with $Q$ classes, SBM and MMSB were run with $Q+1$ classes and
we identified the class of outliers. In practice, this can easily be
done since this class contains most of the vertices of the network
having low output and input degrees.

The code implementing all the experiments is available upon request.

\subsection{Simulations} \label{subsecsimul}

\begin{figure}

\includegraphics{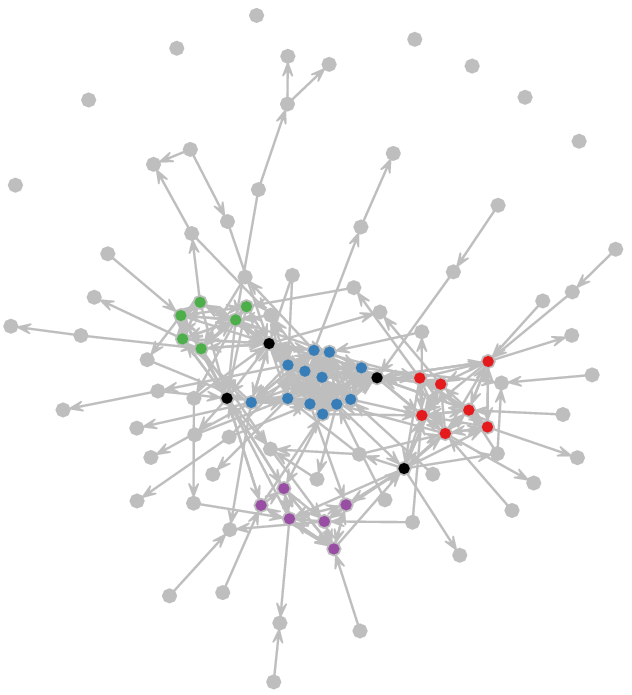}

\caption{Example of a network with community structures. Overlaps are
represented in black and outliers in gray.}
\label{figaffi}
\end{figure}

In this set of experiments we generated two types of networks using
the OSBM generative model. In Section \ref{subsectionAffi} we
sampled networks with community structures (Figure \ref{figaffi}),
where vertices of a
community are mostly connected to vertices of the same community. To
limit the number of free parameters, we considered the $Q \times Q$
real matrix $\mathbf{W}$:
%
\begin{eqnarray}
\mathbf{W}=
\pmatrix{
\bolds{\lambda} & -\varepsilon& \dots& -\varepsilon\cr
-\varepsilon& \bolds{\lambda} & & \vdots\cr
\vdots& & \ddots& -\varepsilon\cr
-\varepsilon& \dots& -\varepsilon& \bolds{\lambda}
}.
\end{eqnarray}
%

\begin{figure}[t]

\includegraphics{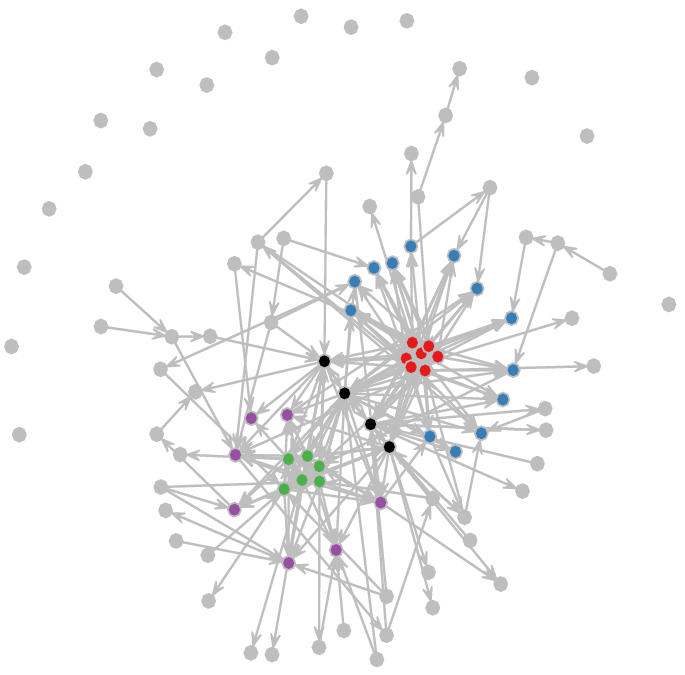}

\caption{Example of a network with community structures and stars.
Overlaps are represented in black and outliers in gray.}
\label{figgraph2}
\end{figure}

In Section \ref{subsectionAffiAndHubs} we generated networks with
more complex topologies, using the matrix $\mathbf{W}$:
%
\begin{eqnarray}
\mathbf{W}=
\pmatrix{
\bolds{\lambda} & \bolds{\lambda} & -\varepsilon& \dots
& \dots& \dots& -\varepsilon\cr
-\bolds{\varepsilon} & -\bolds{\lambda} & -\varepsilon&
\dots& \dots& \dots& \vdots\cr
\vdots& -\varepsilon& \bolds{\lambda} & \bolds{\lambda} &
-\varepsilon& \dots& \vdots\cr
\vdots& \vdots& -\bolds{\varepsilon} & -\bolds{\lambda} &
-\varepsilon& \dots& \vdots\cr
\vdots& \vdots& \vdots& -\varepsilon& \ddots&-\varepsilon& -\varepsilon&
\cr
\vdots& \vdots& \vdots& \vdots& -\varepsilon& \bolds{\lambda} &
\bolds{\lambda} \cr
-\varepsilon& \dots& \dots& \dots& \dots&-\bolds{\varepsilon} &
-\bolds{\lambda}
}.
\end{eqnarray}
In these networks, if class $i$ is a community and has therefore a high
\emph{intra} connection probability, then its vertices also highly connect
to vertices of class $i+1$ which itself has a low \emph{intra}
connection probability. Such star patterns (Figure \ref{figgraph2})
often appear in transcription
networks, as shown in Section \ref{sectionyeast}, and protein--protein
interaction networks.

For these two sets of experiments, we used the $Q$-dimensional real
vectors $\mathbf{U}$ and $\mathbf{V}$:
%
\begin{eqnarray}
\mathbf{U}= \mathbf{V}=
\pmatrix{
\varepsilon& \dots& \varepsilon
}
,
\end{eqnarray}
and we set $Q=4$, $\lambda=4$, $\varepsilon=1$, $W^{*}=-5.5$. Moreover,
for the vector $\bolds{\alpha}$ of class probabilities, we set
$\alpha
_{q}=0.25, \forall q \in\{1,\dots,Q\}$. We
generated $100$ networks with $N=100$ vertices and for each of these
networks, we
clustered the vertices using CFinder, SBM, MMSB and OSBM. Finally, we
used a criterion similar to the one proposed by
\citet{ProceedingsHeller2007}; \citet{ProceedingsHeller2008} to compare the
true $\mathbf{Z}$ and the estimated $\hat{\mathbf{Z}}$ clustering
matrices. Thus, for
each network and each method, we computed the $L_{2}$
distance $d (\mathbf{P}, \hat{\mathbf{P}})$ where $\mathbf
{P}=\mathbf{Z}\mathbf{Z}^{\top}$ and
$\hat{\mathbf{P}}=\hat{\mathbf{Z}}\hat{\mathbf{Z}}{}^{\top}$.
These two $N\times N$
matrices are invariant to column permutations of
$\mathbf{Z}$ and $\hat{\mathbf{Z}}$ and compute the
number of shared clusters between each pair of vertices of a
network. Therefore, $d (\mathbf{P}, \hat{\mathbf{P}})$ is a good
measure to
determine how well the underlying cluster assignment
structure has been discovered. Since CFinder depends on a parameter
$k$ (size of the cliques), for each simulated network, we ran the
software for various values of
$k$ and selected $\hat{k}$ for which the $L_{2}$ distance was
minimized. Note that this choice of $k$ tends to overestimate the
performances of CFinder compared to the other approaches. Indeed, in
practice, when analyzing a real network, $k$ needs to be estimated (see
Section \ref{subsecblog}), while $\mathbf{P}$ is unknown. OSBM was
run with
$Q$ classes, whereas SBM and MMSB were run with $Q+1$ classes. For both
SBM and MMSB, and each generated network, after having identified the
class of outliers, we set the latent vectors of the corresponding
vertices to zero (null component). The $L_{2}$ distance $d(\mathbf{P},
\hat
{\mathbf{P}})$ was then computed exactly as described previously.

\begin{table}[b]
\tablewidth=190pt
\caption{Comparison of CFinder, SBM, MMSB and OSBM in terms of the
$L_{2}$ distance $d (\mathbf{P}, \hat{\mathbf{P}})$ over the $100$
samples of networks with community structures}\label{tableresultsAffi}
\begin{tabular*}{190pt}{@{\extracolsep{4in minus 4in}}lcccc@{}}
\hline
& \textbf{Mean} & \textbf{Median} & \textbf{Min} & \textbf{Max} \\
\hline
CFinder& \phantom{0}43.53 & \phantom{0}22\phantom{.0} & \textbf{0} & \textbf{203} \\
SBM & 116.46 & 103.3 & \textbf{0} & 321 \\
MMSB & \phantom{0}53.76 & \phantom{0}27.5 & \textbf{0} & 293 \\
OSBM & \phantom{0}\textbf{41.83} & \phantom{00}\textbf{0}\phantom{.0} & \textbf{0} & 258 \\
\hline
\end{tabular*}
\end{table}

\subsubsection{Networks with community structures} \label{subsectionAffi}

\begin{figure}[t]

\includegraphics{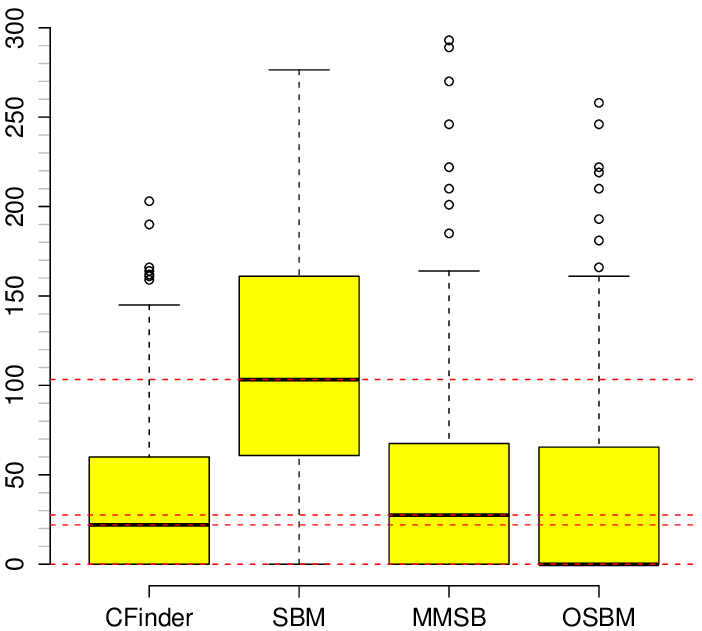}

\caption{$L_{2}$ distance $d (\mathbf{P}, \hat{\mathbf{P}})$ over
the $100$
samples of networks with community structures, for CFinder, SBM, MMSB
and OSBM. Measures how well the underlying
cluster assignment structure has been retrieved.}
\label{figresultsAffi}
\end{figure}

The results that we obtained are presented in Table
\ref{tableresultsAffi} and in Figure
\ref{figresultsAffi}. We can observe that CFinder, MMSB and OSBM
lead to very accurate estimates $\hat{\mathbf{Z}}$ of the true clustering
matrix $\mathbf{Z}$. For most networks, they retrieve the clusters and
overlaps perfectly, although CFinder and MMSB appear to be slightly
biased. Indeed, while the median of the $L_{2}$ distance $d (\mathbf{P},
\hat{\mathbf{P}})$ over the $100$ samples is null for OSBM, it is
equal to
$22$ for CFinder and $27.5$ for MMSB. Since CFinder is an algorithmic
approach, and not a
probabilistic model, it does not classify a vertex $v_{i}$ if it does
not belong to
any $k$-cliques of a $k$-clique community. Conversely, OSBM is more
flexible and can take the random nature of the network into
account. Indeed, the edges are assumed to be drawn randomly, and,
given each pair of vertices, OSBM deciphers whether or not they are
likely to belong to the same class, depending on their connection
profiles.\vadjust{\goodbreak} Therefore, OSBM can predict that $v_{i}$ belongs to a class
$q,$ although it does not belong to any $k$-cliques. Overall, we found
that MMSB retrieves the clusters well but often misclassifies some of
the overlaps. Thus, if a given vertex belongs to several clusters, it
tends to be classified by MMSB into only one of them. Nevertheless, the
results clearly illustrate that MMSB improves over SBM, which cannot
retrieve any of the overlapping clusters. It should also be
noted that CFinder has fewer outliers (Figure \ref{figresultsAffi}) than
MMSB and OSBM and appears to be slightly more stable when looking for
overlapping community structures in networks.

\subsubsection{Networks with community structures and stars}\label{subsectionAffiAndHubs}

\begin{table}[t]
\tablewidth=200pt
\caption{Comparison of CFinder, SBM, MMSB and OSBM in terms of the
$L_{2}$ distance $d (\mathbf{P}, \hat{\mathbf{P}})$ over the $100$
samples of networks with community structures and stars}\label{tableresults2}
\begin{tabular*}{200pt}{@{\extracolsep{4in minus 4in}}lcccc@{}}
\hline
& \textbf{Mean} & \textbf{Median} & \textbf{Min} & \textbf{Max} \\
\hline
CFinder& 362.07 & 354.5\phantom{0} & 181\phantom{.00} & 567\phantom{.00}\\
SBM & 134.68 & 118.87 & \phantom{0}15.14 & 352.09 \\
MMSB & 119.01 & \phantom{0}98.5\phantom{0} & \phantom{00}\textbf{0}\phantom{.00} & 367\phantom{.00} \\
OSBM & \phantom{0}\textbf{77}\phantom{.00} & \phantom{0}\textbf{43}\phantom{.00} & \phantom{00}\textbf{0}\phantom{.00} & \textbf{328}\phantom{.00} \\
\hline
\end{tabular*}
\end{table}

\begin{figure}[b]

\includegraphics{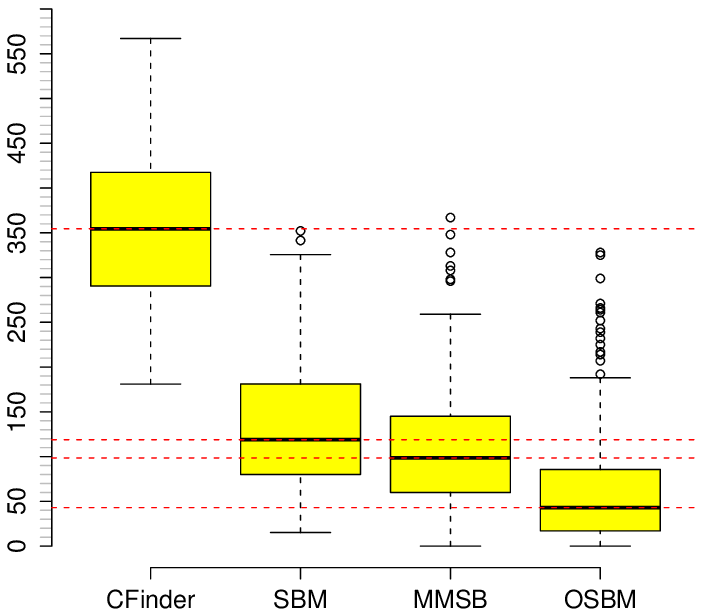}

\caption{$L_{2}$ distance $d (\mathbf{P}, \hat{\mathbf{P}})$ over
the $100$
samples of networks with community structures and stars, for CFinder,
SBM, MMSB and OSBM. Measures how well the underlying
cluster assignment structure has been retrieved.}
\label{figresults2}
\end{figure}

In this set of experiments we considered networks with more complex
topologies. As shown, in Table \ref{tableresults2} and in Figure
\ref{figresults2}, the results of CFinder dramatically degrade while
those of OSBM remain more stable. Indeed, the median of the $L_{2}$
distances $d (\mathbf{P}, \hat{\mathbf{P}})$ over the $100$ samples
is equal to $43$
for OSBM, while it is equal to $354.5$ for CFinder. This can be easily
explained since CFinder only looks for community structures of
adjacent $k$-cliques, and cannot retrieve classes with low
\emph{intra} connection probabilities. Conversely, OSBM uses a $Q
\times Q$ real matrix $\mathbf{W}$ and two real vectors $\mathbf{U}$
and $\mathbf{V}$ of size
$Q$ to model the \emph{intra} and \emph{inter} connection
probabilities. No assumption is made on these matrix and vectors such
that OSBM can take heterogeneous and complex topologies into account.
As for CFinder, the results of MMSB degrade, although they remain
better than SBM.
As for the previous Section, MMSB retrieves the clusters well but
misclassifies the overlaps more frequently when considering networks
with community structures and stars.

\subsection{French political blogosphere} \label{subsecblog}

\begin{figure}[b]

\includegraphics{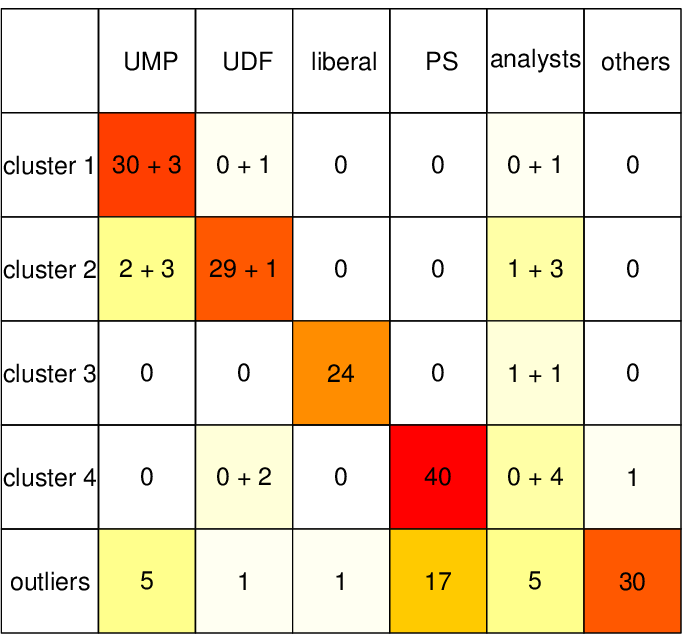}

\caption{Classification of the blogs into $Q=4$ clusters using OSBM.
The entry $(i, j)$ of the matrix describes the number of blogs
associated to the $j$th political party (column) and classified into
cluster $i$ (row). Each entry distinguishes blogs which belong to a
unique cluster from overlaps (single membership blogs $+$ overlaps).
The last row corresponds to the null component.}
\label{figresultsOSBM}
\end{figure}

We consider the French political blogosphere network and we focus on a
subset of 196 vertices connected by 2864 edges. The data consists of a
single day snapshot of political blogs automatically extracted on the 14th
of October 2006 and manually classified by the ``Observatoire
Pr\'{e}sidentielle project'' [\citet{ArticleZanghi2008}]. Nodes
correspond to hostnames and there is
an edge between two nodes if there is a known hyperlink from one
hostname to another. The four main political parties which are present
in the data set are the UMP (french ``republican''), UDF (``moderate''
party), liberal party (supporters of economic-liberalism) and PS
(french ``democrat''). Therefore, we applied our algorithm with $Q=4$
clusters and
we obtained the results presented in Figure \ref{figresultsOSBM} and
Table \ref{tableestimatedW}.

\begin{table}[t]
\tablewidth=250pt
\caption{The estimated $\widetilde{\mathbf{W}}$ matrix for the
classification of
the blogs into $Q=4$ clusters using OSBM. The $4\times4$ matrix on the
top left-hand side represents the $\mathbf{W}$ matrix, while the
vectors on
the top right-hand side and bottom left-hand side represent the vectors
$\mathbf{U}$ and $\mathbf{V}^{\top}$ respectively. The
remaining term
corresponds to the bias. The diagonal of $\mathbf{W}$ indicates that blogs
have a heavy tendency to connect to blogs of the same class. Blogs of
cluster 1 (UMP) have also a positive tendency to connect to blogs of
clusters 2 (UDF) and 3 (liberal party). Conversely, blogs of cluster 4
(PS), representing the left wing, are more isolated in the network}
\begin{tabular*}{250pt}{@{\extracolsep{4in minus 4in}}lccp{29pt}c@{}}
\hline
\phantom{$-$}3.89 & \phantom{$-$}0.17 & \phantom{$-$}0.54 & $-$0.70 &$-$0.70 \\
\phantom{$-$}0.17 & \phantom{$-$}2.47 & $-$0.40 & $-$0.84 & \phantom{$-$}0.40 \\
\phantom{$-$}0.55 & $-$0.40 & \phantom{$-$}4.43 & $-$0.85 & $-$0.38 \\
$-$0.70 & $-$0.84 & $-$0.85 & \multicolumn{1}{c}{\phantom{$-$}1.66\phantom{0.}} & \phantom{$-$}0.87 \\[6pt]
$-$0.70 & \phantom{$-$}0.40 & $-$0.38 & \multicolumn{1}{c}{\phantom{$-$}0.87\phantom{0.}} & $-$3.60 \\
\hline
\end{tabular*}
\label{tableestimatedW}
\end{table}

First, we notice that the clusters we found are highly homogeneous and
correspond to the well-known political parties. Thus, cluster 1 contains
35 blogs among which 33 are associated to UMP, while cluster 2 contains
39 blogs among which 30
are related to UDF. Similarly, it follows that
cluster 3 corresponds to the liberal party and cluster 4 to PS. We
found nine overlaps. Thus, three blogs associated to UMP belong to both
clusters 1 (UMP) and 2 (UDF). This is a result we expected since these
two political parties are known to have some relational ties. Moreover,
a blog
associated to UDF belongs to both clusters 1 (UMP) and 4 (PS), while
another UDF blog belongs to clusters 2 (UDF) and 4 (PS). This can
be easily understood since UDF is a moderate party. Therefore, it is
not surprising to find UDF blogs with links with the two biggest
political parties in France, representing the left and right
wings. Very interestingly, among the nine overlaps we found, four of
them correspond to blogs of political analysts. Thus, a blog overlaps
clusters 1~(UMP) and 4 (PS). Another one overlaps clusters 2 (UDF), 3 (liberal
party) and 4~(PS). Finally, the two last blogs of political analysts
overlap clusters 2 (UDF) and 4 (PS).

We ran CFinder and we used the criterion [\citet{ArticlePalla2005}]
they proposed to select
$k$ (see Section \ref{secIntroduction}). Thus, we ran the software
for various values of $k$ and we found $\hat{k}=7$. Lower
values lead to giant components which smear the details of the network.
Conversely, for higher values, the
communities start disintegrating. Using $\hat{k}$, we uncovered $11$
clusters which correspond to sub-clusters of the clusters we found
using OSBM. For instance, cluster 3 (liberal party) was split into two
clusters, whereas cluster 4 (PS) was split into three. Indeed, while
OSBM predicted that the connection profiles of these sub-clusters
were very similar and therefore should be merged, CFinder could not
uncover any $k$-clique community, that is, a union of \emph{fully} connected
sub-graphs of size $k$, containing these sub-clusters. Note that using
CFinder, we retrieved the overlaps uncovered by our algorithm. CFinder
did not classify 95 blogs.

\begin{figure}[t]

\includegraphics{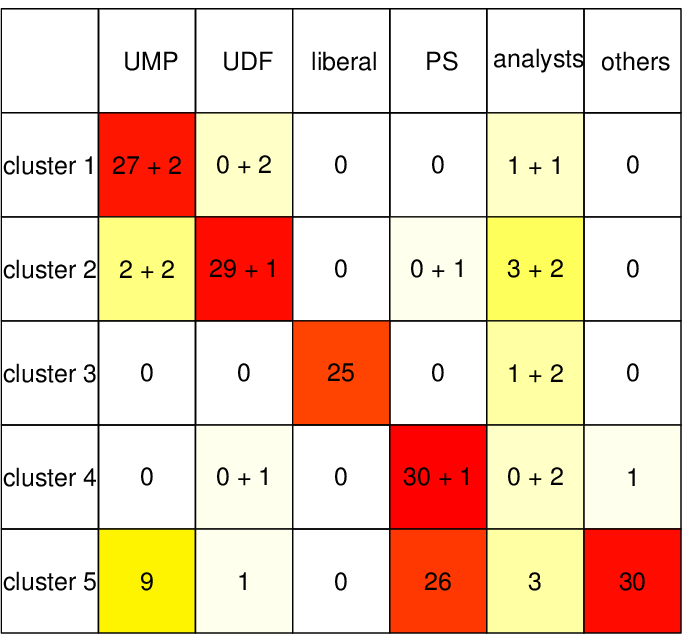}

\caption{Classification of the blogs into $Q=5$ clusters using MMSB.
The entry $(i, j)$ of the matrix describes the number of blogs
associated to the $j$th political party (column) and classified into
cluster $i$ (row). Each entry distinguishes blogs which belong to a
unique cluster from overlaps (single membership blogs $+$ overlaps).
Cluster 5 corresponds to the class of outliers.}
\label{figresultsMMSB}
\end{figure}

\begin{figure}[t]

\includegraphics{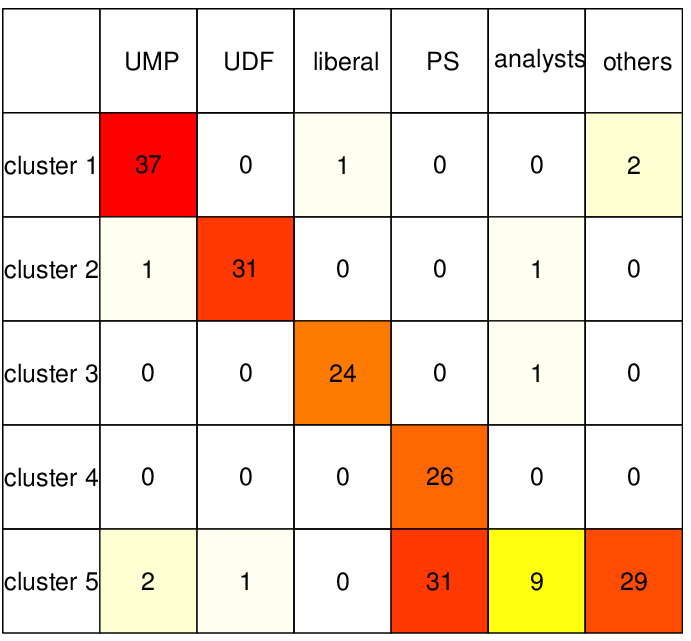}

\caption{Classification of the blogs into $Q=5$ clusters using SBM.
The entry $(i, j)$ of the matrix describes the number of blogs
associated to the $j$th political party (column) and classified into
cluster~$i$ (row). Cluster 5 corresponds to the class of outliers.}
\label{figresultsSBM}
\end{figure}

We also clustered the blogs of the network using MMSB and SBM. As
previously, for both models, we used $Q+1$ clusters and we identified
the class of outliers. The results of MMSB are presented in Figure \ref
{figresultsMMSB}. Overall, we can notice that MMSB led to similar
clusters as OSBM, although cluster 4 is less homogeneous in MMSB than
in OSBM. 
We found eight overlaps using MMSB and we emphasize that five of them
correspond exactly to the one found with our approach. Thus, the model
retrieved two among the three UMP blogs overlapping clusters~1~(UMP)
and 2 (UDF). Moreover, MMSB uncovered the UDF blog overlapping clusters~1~(UMP) and 4 (PS), as well as the blog of political analysts
overlapping clusters 2 (UDF), 3 (liberal party) and 4 (PS). It also
retrieved the blog of political analysts overlapping clusters~1~(UMP)
and 4 (PS).
Finally, the results of SBM are presented in Figure \ref
{figresultsSBM}. Again, the clusters found by this approach are very
similar to the one uncovered by OSBM. However, because SBM does not
allow each vertex to belong to multiple clusters, it misses a lot of
information in the network. In particular, while some of the blogs of
political analysts are viewed as overlaps by OSBM, because of their
relational ties with the different political parties, they are all
classified into a single cluster by SBM.

\begin{table}[b]
\caption{Classification of the operons into $Q=6$ clusters. Operons
in bold belong to multiple clusters}
\label{Tableresults}
\begin{tabular*}{\tablewidth}{@{\extracolsep{4in minus 4in}}lcl@{}}
\hline
\textbf{Cluster} & \textbf{Size} & \multicolumn{1}{c@{}}{\textbf{Operons}} \\
\hline
1 & \phantom{0}2 & \multicolumn{1}{l}{STE12 TEC1} \\[3pt]
2 & 33 &YBR070C MID2 YEL033W SRD1 TSL1 RTS2
PRM5 YNL051W PST1\\
&&YJL142C \textbf{SSA4}
YGR149W SPO12 YNL159C SFP1 YHR156C YPS1 \\
&&YPL114W HTB2
MPT5 SRL1 DHH1
\textbf{TKL2} PGU1 YHL021C RTA1\\
&&WSC2 GAT4 YJL017W TOS11 YLR414C
BNI5 YDL222C
\\[3pt]
3 & \phantom{0}2 & \multicolumn{1}{l}{MSN4 MSN2} \\[3pt]
4 & 32 & CPH1 \textbf{TKL2} HSP12 SPS100 MDJ1 GRX1
SSA3 ALD2 GDH3 GRE3\\
&& HOR2
ALD3 SOD2 ARA1 HSP42 YNL077W \textbf{HSP78} GLK1 DOG2\\
&& HXK1 RAS2
\textbf{CTT1} HSP26 TPS1 TTR1 HSP104 GLO1 \textbf{SSA4}
PNC1\\
&& MTC2 YGR086C
\textbf{PGM2}
\\[3pt]
5 & \phantom{0}2 & \multicolumn{1}{l}{YAP1 SKN7} \\[3pt]
6 & 19 &YMR318C \textbf{CTT1} TSA1 CYS3 ZWF1
HSP82 TRX2 GRE2 SOD1 AHP1\\
&&
YNL134C \textbf{HSP78} CCP1 TAL1 DAK1 YDR453C TRR1
LYS20
\textbf{PGM2}\\
\hline
\end{tabular*}
\end{table}


\subsection{Saccharomyces cerevisiae transcription network}
\label{sectionyeast}

We consider the yeast transcriptional regulatory network described in
\citet{ArticleMilo2002} and we focus on a subset of 197 vertices
connected by 303 edges. Nodes of the network correspond to operons, and
two operons are linked if one operon encodes a transcriptional factor
that directly regulates the other operon. The network is made of three
regulation patterns, each one of them having its own regulators and
regulated operons. Therefore, using $Q=6$ clusters, we applied our
algorithm and
we obtained the results in Table \ref{Tableresults}.

First, we notice that clusters 1, 3 and 5 contain only two
operons each. These operons correspond to hubs which regulate
respectively the nodes of clusters 2, 4 and 6, all having a very low
\emph{intra} connection probability. To analyze our results, we used
GOToolBox [\citet{ArticleMartin2004}] on each cluster. This software
aims at identifying statistically over-represented
terms of the Gene Ontology (GO) in a gene data set. We found that the
clusters correspond to well-known
biological functions. Thus, the
nodes of cluster 2 are regulated by STE12 and TEC1 which are both
involved in the response to glucose
limitation, nitrogen limitation and abundant fermentable carbon
source. Similarly, MSN4 and MSN2 regulate the nodes of cluster 4 in
response to different stress such as freezing, hydrostatic pressure
and heat acclimation.
Finally, the nodes of cluster 6 are regulated by YAP1 and SKN7 in the
presence of oxygen stimulus. Our algorithm was able to uncover two
overlapping clusters
(operons in bold in Table \ref{Tableresults}). Interestingly,
contrary to the other operons of clusters 2, 4 and 6, which are all
regulated by operons of a single cluster (clusters 1, 3 or 5), these
overlaps correspond to co-regulated operons. Thus, SSA4 and TKL2 belong
to clusters 2 and
4 since they are co-regulated by (STE12, TEC1) and (MSN4 and
MSN2). Moreover, HSP78, CTT1 and PGM2 belong to clusters 4 and 6 since
they are co-regulated by (MSN4, MSN2) and (YAP1, SKN7). It should also
be noted that OSBM did not classify 112 operons which all have very low
output and input degrees.

Because the network is sparse, we obtained very poor results with
CFinder. Indeed, the network contains only one $3$-clique and no
$k$-clique for $k>3$. Therefore, for $k=2$, all the operons were
classified into a single cluster and no biological information could
be retrieved. For $k=3$, only three operons were classified into a
single class and for $k>3$ no operon was classified.

As previously, we ran MMSB and SMB with $Q+1$ clusters and we
identified the class of outliers. Both approaches retrieved the six
clusters found by OSBM. However, we emphasize that, contrary to the
political blogoshpere network, MMSB did not uncover any overlap in the
yeast transcriptional regulatory network.

As in Section \ref{subsecsimul}, these results clearly illustrate the
capacity of OSBM to retrieve overlapping clusters in networks with
complex topological structures. In particular, in situations where
networks are not made of community structures, while the results of
CFinder dramatically degrade or cannot even be interpreted, OSBM
seems particularly promising.

\section{Conclusion}

In this paper we proposed a new random graph model, the Overlapping
Stochastic Block Model, which can be used to retrieve overlapping
clusters in networks. We used global and local variational techniques
to obtain a tractable lower bound of the observed log-likelihood and
we defined an EM like procedure which optimizes the model
parameters in turn. We showed that the model is identifiable within
classes of equivalence and we illustrated the efficiency of our
approach compared to other methods, using simulated data and real networks.
Since no assumption is made on the matrix $\mathbf{W}$ and vectors
$\mathbf{U}$ and
$\mathbf{V}$ used to characterize the connection probabilities, the
model can
take very different topological structures into account and seems
particularly promising for the analysis of networks. In the experiment
section we set the number $Q$ of classes using {a priori}
information we had about the networks. However, in future
works, we believe it is crucial to develop a model selection criterion
to estimate the number of classes automatically from the topology. We
will also investigate introducing
some priors over the model parameters to work in a full Bayesian framework.

\section*{Acknowledgment}

The authors would like to thank C. Matias for her helpful remarks and
suggestions for the proof on model identifiability.

\begin{supplement}[id=appA]
\sname{Supplement}
\stitle{Appendix}
\slink[doi]{10.1214/10-AOAS382SUPP} 
\slink[url]{http://lib.stat.cmu.edu/aoas/382/supplement.pdf}
\sdatatype{.pdf}
\sdescription{Describe how global and local variational techniques can
be used to obtain a tractable lower bound. Introduce the optimization
equations for the inference procedure.}
\end{supplement}


%


\printaddresses

\end{document}